\documentclass{article}

\usepackage{arxiv}

\usepackage[utf8]{inputenc} 
\usepackage[T1]{fontenc}    
\usepackage{hyperref}       
\usepackage{url}            
\usepackage{booktabs}       
\usepackage{amsfonts}       
\usepackage{nicefrac}       
\usepackage{microtype}      
\usepackage{lipsum}		
\usepackage{graphicx}
\usepackage{natbib}
\usepackage{doi}

\usepackage{amsmath}
\usepackage{float}
\usepackage{mathtools}

\title{Information theoretic analysis of computational models as a tool to understand the neural basis of behaviors}

\author{Madhavun Candadai \\
	Cognitive Science Program\\
	Indiana University\\
	Bloomington, IN 47405 \\
	\texttt{madvncv [at] gmail.com} \\
}

\renewcommand{\shorttitle}{\textit{arXiv} Template}

\hypersetup{
pdftitle={Information theory + computational models in Neuroscience},
pdfauthor={Madhavun Candadai},
pdfkeywords={Computational models, Information theory, Embodied cognition, Neural basis of behaviors, Neural Networks},
}

\begin{document}
\maketitle

\begin{abstract}
One of the greatest research challenges of this century is to understand the neural basis for how behavior emerges in brain-body-environment systems. To this end, research has flourished along several directions but have predominantly focused on the brain. While there is in an increasing acceptance and focus on including the body and environment in studying the neural basis of behavior, animal researchers are often limited by technology or tools. Computational models provide an alternative framework within which one can study model systems where ground-truth can be measured and interfered with. These models act as a hypothesis generation framework that would in turn guide experimentation. Furthermore, the ability to intervene as we please, allows us to conduct in-depth analysis of these models in a way that cannot be performed in natural systems. For this purpose, information theory is emerging as a powerful tool that can provide insights into the operation of these brain-body-environment models. In this work, I provide an introduction, a review and discussion to make a case for how information theoretic analysis of computational models is a potent research methodology to help us better understand the neural basis of behavior.
\end{abstract}


\section{Introduction}
The number of questions that we as scientists ask about natural systems, and the number of hours it would take to perform the experiments required to answer those questions far outnumber the number of hours we collectively have. This is true just for the questions that we know to ask at this moment in time. A hallmark of seminal work in any scientific discipline is its ability to open up the possibilities for new questions to be asked. Furthermore, the solutions that evolution found to problems that natural systems face, significantly dwarfs our imagination and problem-solving capabilities. Thus, the quest for understanding natural systems is a challenging feat that scientists take on fully aware that their best-case-scenario would be to make a tiny dent in the otherwise vast ever-expanding envelope of scientific progress. Rest assured, this bleak opening statement about the limited capacities of the human intellect, while humbling, is a segue into the benefits of taking a computational modeling approach to science. 

Models of natural systems are often considered to be built as a tool to generate predictions about the system under study, like a model of thunderstorms used to predict its movement trajectory. However, the objective of building models goes beyond making predictions~\citep{epstein2008}. Models serve to: provide an explanatory account of observed phenomena, guide experimental design and data collection based on hypothesis generated from analyzing the model, suggest analogies by demonstrating parallels between different natural systems, provide existence-proofs for mechanisms underlying behavior of natural systems thus raising new scientific questions, explore the hypothesis space at a comparatively low cost compared to experimental settings, and finally, act as idealized test beds for novel analytical methods since the ground-truth knowledge is available about these systems. Importantly, the scientific approach to modeling involves building ``opaque'' models constrained only by known details, only to then be analyzed later to reveal the internal workings. As a result of this, modelers naturally adopt the "embracing multiple hypotheses" approach, an approach that arguably leads to better and less-biased science~\citep{chamberlin1890method}. When integrated into a modeling-experiment cycle, computational models can continuously inform future experimental design while getting updated based on results from past experiments~\citep{alexander2015reciprocal}. This virtuous cycle of experiments and models informing each other has been proven to provide a structured way to explore the intractable hypothesis space for understanding natural systems~\citep{izquierdo2019role}. One of the greatest scientific challenges of this century is to understand how the brain produces behavior and computational models stand to make a great difference. 

Approaches to analyzing and understanding neural network operation in the context of behavior requires that it captures linear as well as non-linear interactions between the different components: environment-neural network and neuron-neuron. Several methodologies have been proposed to capture specific features of neural activity such as connectivity, effective dimensionality, encoding, and decoding. In this regard, information theory has emerged as one of the main theoretical frameworks that allow us to study neural information processing at all spatial and temporal scales. It provides tools to quantify these relationships in a way that is invariant to the scale of the system and allows comparison across systems. For instance, information theory allows us to answer questions such as, "how much does knowing the value of one component of a system reduces uncertainty about another component?" (a generalized correlation, if you will), "how much information about a variable of interest is transferred from one part of the system to another?" (enables studying the flow of information about a task relevant variable), "amongst these different sources that are feeding information into this component, what is amount of unique information that each source provides, how much is redundant and how much is available due to combined knowledge of receiving information from each source?" and so on. This paper aims to introduce the different tools that allow exploring such questions, provides explanation on practically estimating these quantities, and examples of how they have been used in the other work.

In this work I discuss and justify an approach to understanding the neural basis of behaviors that involves (a) building computational models of brain-body-environment systems (b) optimization of these models to meet certain criteria and (c) in-depth analysis of the models to develop hypotheses and proofs-of-existence for how natural systems might be operating. This paper is organized as follows: First, I discuss the benefits of employing computational models in science, specifically in Neuroscience with examples of previous work that demonstrate how they have helped advance our understanding of neural network operation; second, I discuss the different components of building computational models such as choosing a neural network model and optimization algorithm; third, I explain how the tools of information theory have been used to understand neural information processing; and finally, I present how research can combine these above mentioned methodologies to study the neural basis of behaviors.

\section{Computational models in Neuroscience}
Thought-experiments have long been drivers of scientific progress by identifying areas for experimentalists to focus on, based on careful observation of known facts of a phenomena. Simulation models can serve as a mathematically grounded replacement to thought-experiments~\citep{di2000simulation, nersessian2012computational}, with the potential to go beyond the biases that we as humans bring into thought experiments. 

Computational models (or models of any kind) are tractable versions of the complex natural system that we aim to understand. Justifiable simplifications in combination with powerful computing systems have enabled the construction of a myriad of computational models in Neuroscience ranging from models of individual neurons to integrated brain-body-environment models. Some of the apparent practical benefits of computational models are as follows: first, unlike experimental in settings computational models are fully observable - all variables are completely accessible at all times of the simulation, and beyond being observable, all variables in the system are manipulable and enable ablation as well as stimulation tests in any part of the system. However, like the statistician George E. P. Box eloquently stated, "All models are wrong but some are useful". Computational models are idealized abstractions of the system of interest and therefore necessarily make simplifying assumptions in their design. Consequently, insights obtained from models unlike those from experiments do not serve as definitive proof but instead provide hypotheses and existence proofs that would then need to be tested. 

\cite{cohen2004mathematics} delineated six types of questions that are asked by scientists: How did it begin? (Origins), How is it built? (Structures), What is it for? (Functions), How does it work? (Mechanisms), What goes wrong? (Pathologies), and How is it fixed? (Repairs). Computational models are useful in Neuroscience to the extent that they help answer at least one of these questions. In this section, I outline examples of computational models being applied to each of these questions within the domain of Neuroscience. I only point out examples to provide a proof-of-existence; this is by no means an exhaustive list. 

\subsection{How did it begin? (Origins)}
New phenomena originate in natural systems at several time-scales, from the first time it appears in any species such as the first occurrence of a neuron in evolutionary time~\citep{villegas2000origin}; to the first occurrence in an individual over developmental time scales such as the origin of cortical interneurons hypothesized to be in cortical subventricular zone~\citep{wonders2006origin}; to the origin of specialized neural dynamics during on going behavior such as what causes different types of oscillations to begin in the same neural circuit of the crab stomatogastric ganglion~\citep{marder2007understanding}. An example of how computational models have been used to answer questions of origin is the work by \cite{krishnan2018origin}, that provided a hypothesis for the origin of spontaneous resting state dynamics observed in fMRI, EEG and local field potential recordings. Using a computational model of intra- and extracellular $K^{+}$ and $Na^{+}$ ion flow across the brain network based on the CoCoMac connectivity data, they were able to reproduce the infra-slow fluctuations in brain activity. Thus, a computational model has helped hypothesize that the origin of spontaneous neural dynamics could be due to dynamics of ion flow mediated by neuronal and glial activity. 

\subsection{How is it built? (Structures)}
Sometimes, there is only partial knowledge of the different components of a natural system and a complete understanding of its function requires that the gaps are filled. For instance, even in \textit{C. elegans} which has been nearly completely genetically, developmentally and neurologically mapped, while the entire connectome of the ~302 neuron network is known, the polarity of the connections (inhibitory vs excitatory) are not known. This is particularly true in larger animals where the micro-connectome is not fully known. An example of where computational models can be utilized to advance our understanding of such systems is the work by \cite{real2017neural}, where they optimized a model to match observed functional characteristics constrained by known structural properties to make predictions about the unknown structural attributes of the vertebrate retinal ganglion. Specifically, they started with an existing model made with known anatomical details and built four models that incorporated different anatomical characteristics in sequence: first, the addition of a non-linear pooling layer; second, a feedback loop around the non-linear pooling layer; third, another feedback loop around but this time around a different downstream layer that feedback of that matched firing rate of ganglion cells; and finally, a delay in processing before the pooling layer. These four modifications were chosen based on the expectation of matching increasingly more functional characteristics that were observed in the retinal ganglion. They showed that these four modifications when applied in a cascaded fashion performed progressively better in explaining the variance in activity of the ganglion cells. Importantly, the model was re-discovered known attributes of ganglion cells such as center-surround receptive fields at appropriate scales in each layer in accordance with observed experimental data. Furthermore, they conducted experiments to test if the nature of the receptive fields formed in the model matched with what can be observed biologically and found that they indeed matched. 
Thus, this study is an example for how models can help answer the question of how a neural network is built to perform a specific function.

\subsection{What is it for? (Functions)}
A sub-problem in developing a thorough understanding of a natural system is understanding the role played by its different component systems. Experimental results often provide an incomplete account of the role of a component because experiments are often geared towards testing one specific hypothesis. As a result, different experiments might provide seemingly conflicting accounts that will require reconciliation. The computational modelling example for answering questions of function comes from the work of \cite{alexander2010computational} regarding the role of Anterior Cingulate Cortex (ACC) in reinforcement learning. Their Predicted Response-Outcome model involved representing a vector-valued prediction of a sequence of future states of the prefrontal cortex given the current state and action. Further, in accordance with existing experimental data they reinterpreted the polarity of the prediction error to mean positive surprise (events that were not predicted but occurred) and negative surprise (events that were predicted but did not occur). Crucially, according to this model, it does not matter if a predicted state is desirable or undesirable, but it only matters whether the outcome was predicted or not. This generalized model of the role of ACC as providing ``valence-neutral prediction error'' provided a unified account of seemingly disparate experimental observations that was thought to be impossible to reconcile. Additionally, the model makes predictions for action selection based on a combination of learned state-response associations and response-outcome associations. Thus, computational approaches have enabled identification of the functional role of the ACC thereby helping answer what is it for. 

\subsection{How does it work? (Mechanisms)}
In addition to understanding the role of individual components, it is crucial to understand how the different components interact to produce the observable phenomenon. This is another context where computational models could help generate testable hypotheses. Such an example is the work of \cite{olivares2018potential} which provided existence-proof of purely central pattern generator driven locomotion in \textit{C. elegans}. They built a neural network model of the repeating neural unit of the segmented worm constrained by the known connectome, and optimized the unknown parameters such that the model generated neural activity that matched dynamics expected during locomotion. Analyzing an ensemble of models that performed similarly revealed multiple ventral nerve cord architectures that could result in oscillatory dynamics suitable for locomotion. The most frequently found solution in the ensemble involved a dorsal oscillatory circuit that drove the ventral out-of-phase oscillations thereby producing the requisite alternating dynamics for locomotion. Following this, experimental studies have provided support for the possibility of intrinsic oscillations in the ventral nerve cord~\citep{xu2018descending, fouad2018distributed, gao2018excitatory}. Thus, computational models can help explore the potential space of mechanisms that can produce a behavior. 

\subsection{What goes wrong? (Pathologies)}
The first step in treating any pathology is understanding its cause. To this end, several hypotheses might be proposed that would result in the observed phenotypic abnormality. For instance, schizophrenia was theorized to be caused by disruptions in the interactions between dopamine and the prefrontal cortex thus disabling the individual by not being able to hold or update information about environmental context. \cite{braver1999cognition} built a computational model that incorporated a noisy dopamine signal and the subsequent improper modulation of information processing in the prefrontal cortex. This model replicated the expected behavioral aberrations in a continuous performance test that has been shown to capture critical aspects of cognitive control that are affected in schizophrenic individuals. In fact, such approaches have led to development of the sub-field devoted to applying computational approaches to mental disorders: computational psychiatry~\citep{wang2014computational}. Thus, a computational model has provided further support for a theory by concretizing it in a mathematical framework. 

\subsection{How is it fixed? (Repairs)}
Besides being subject to pathological interruptions, natural systems are capable of recovering and self-repair in a variety of ways. Computational models can not only model the pathology in neural networks but also the process that following a pathological interference. One such example is the work of \cite{naeem2015role} who built a computational model of the self-repair process in neural circuits mediated by astrocytes. They proposed a new learning rule that mediated interaction between astrocytes and neurons to reproduce the self-repair process. While the astrocytes were coupled to neurons they also implemented communication between astrocytes to achieve network level repair. They demonstrated that their proposed learning rule reestablished the firing rates of neurons post failure of synapses in a neural network. Thus, computational models enable study of neural processes that facilitate repair following trauma. 

\subsection{Validating analytical methods}
In addition to the scientific questions listed above, the fact that the ground-truth about operations in computational model is known has been used to demonstrate the validity of analytical methods. For instance, \cite{ito2011extending} proposed an extension to transfer entropy that will allow the detection of transfers at different delays, and demonstrated its capability by testing it on spiking neural network model with a known connectivity pattern and range of synaptic delays. While the existing approach that only accounted for delays of one time step captured only 36\% of the true connections in the network, their proposed enhancement captured as much as 73\% of them. Thus, computational models serve as an ideal test bed for evaluating and validating analytical methods before applying them to experimental data.


\subsection{Replicating natural systems}
Inspired by the adaptability and robustness of natural systems, the entire field of Artificial Intelligence aims to replicate natural intelligence in digital systems. To this end, neural network models of various kinds have been developed to mimic biological neural processes. Tremendous progress has been made in the last few decades with the advent of optimization methods such as backpropagation that enable training multi-layer neural networks. With human- and super-human-level performance as the goal, such artificial systems have become masters of strategy games such as chess~\citep{silver2018general}, Go~\citep{silver2017mastering}, as well as Arcade games~\citep{mnih2013playing} and multiplayer video games~\citep{berner2019dota}. From a practical standpoint, such models have been adopted for aiding cancer diagnosis~\citep{levine2019rise}, aggregating information about global events from public crowd-sourced images~\citep{wang2013observing} and so on. While such systems have a long way to go in order to achieve the robustness of natural systems, rapid progress made in the last decade shows promise for the future.

Altogether, a computational modeling approach spans all levels of scientific inquiry from origin to destruction. The examples provided here are one of several in each of those domains. As mentioned previously, computational models provide unique benefits that enable in-depth analysis and manipulability that cannot be achieved in experimental settings. Computational models have a long history of aiding experimentalists, and will continue to develop along side experiments as a resource for generating hypothesis, testing feasibility of theories and validating analytical approaches.

\section{Model optimization approaches}
There is a wide-variety of computational models of neurons and neural networks that can be employed (Appendix A provides a primer on the most widely-used models). Irrespective of the choice of neuron model, they all have parameters that need to be tuned such that the model behaves as desired. Such tuning cannot be performed by hand-designing these parameters. Several optimization methodologies have been proposed for this purpose. Generally speaking, the optimization process involves defining an objective function as a function of the parameters of the model, and identifying the optima on the objective function landscape. Based on the nature of the objective function, model training falls under one of three approaches: supervised, unsupervised and reinforcement learning. Neural network models where the true desired output of the model is known can be optimized using an objective function that minimizes the error between actual and desired outputs. Such a training paradigm is defined as supervised learning. Alternatively, when the ground-truth outputs are not known, under the paradigm of unsupervised learning, objective functions involve inferring statistics from the training data. Finally, reinforcement learning involves optimizing model parameters to maximize the objective function defined as the expected long-term reward in sensorimotor tasks. Based on the objective function, parameter optimization approaches fall under two primary categories: gradient-based and search-based approaches.

Given an objective function, $Q$, and a set of parameters, $\theta$, gradient-based approaches involve estimating the direction in parameter space to move so as to improve the model performance. This is achieved by estimating the derivative of the objective function with respect to the parameters. Consequently, this requires that the objective function for gradient-based optimization is differentiable. Typically, estimating the true gradient of the objective function is intractable. The gradient is often estimated using a random sample of training data leading to stochastic estimates of the true gradient hence called stochastic gradient-descent. For supervised learning the data is a random sample of training data and labels, and for reinforcement learning, gradients are estimated from sensory input, motor action, and reward data points collected during the course of a sensorimotor task. Since this approach is sensitive to learning rate, several variants have been proposed to this approach that modulate the learning rate: adding a momentum term where learning rate is increased when updates progressively happen in the same direction, and decreases otherwise~\citep{rumelhart1986learning}; adaptive gradient, or AdaGrad, where each parameter has its own learning rate that is increased for sparser parameters and decreases learning rate for less sparse parameters~\citep{duchi2011adaptive}; Root Mean Square Propagation, or RMSProp, that also has per-parameter learning rates which are modulated based on past-updates to the parameters~\citep{tieleman2012lecture}; and finally Adaptive Momentum Estimation, or Adam, is an update to RMSProp where learning rate of each parameter is not just updated based on average of past gradients but also variance of past gradients~\citep{kingma2014adam}. This general approach to incrementally updating model parameters can be applied to any objective function that is differentiable. Depending on the task, the objective function can either require minimization of an error, or maximization of returns. 

An alternative to gradient-based approaches to model optimization are search based approaches. These typically involve a population of solutions searching the parameter space in parallel, and exchanging information on their relative performance in some form according to an objective function as described previously. However, since updates to parameters are not based on estimating gradients, these approaches do not require that the objective function be differentiable. Consequently, the objective function can be more interpretable measures of model behavior. The most popular search based approaches are genetic and evolutionary algorithms~\citep{mitchell1998introduction, holland1992adaptation}. These algorithms get their name since they are inspired by the natural evolutionary process of fitness-based selection. The algorithm is inherently a maximization algorithm on a population of solutions, where each individual in the population is represented as a genotype, a point in the N-dimensional parameter space. In its simplest implementation, each step, or generation, involves: first, estimating the performance or fitness of a subset or the entire population (depending on which variant of evolutionary algorithms are employed) using the objective function; second, copying parameters from the highest fitness individuals to the low fitness individuals or recombination; and third, adding random noise to the low fitness individuals, or mutation. This process is repeated until at least one solution of a desired fitness is found, or until a predetermined computational budget has been exhausted. 

Variations to evolutionary approaches typically involve different methods of selecting individuals to recombine: rank-based selection where individuals are ranked based on their fitness and picked with probabilities according to their rank; fitness-proportionate selection where a probability distribution of selecting individuals is constructed proportional to the relative fitness of the different individuals; elitist-selection where a top $X\%$ of high-fitness individuals are selected to be preserved as is for the next generation; and finally, tournament selection where two individuals are picked at random and pit against each other with the high-fitness individual being preserved as is while ``transfecting'' the low-fitness individual that is then mutated. 

Variation to evolutionary approaches also involve multiple recombination strategies: one-point crossover where all genes in a genotype that are on one side of an arbitrarily chosen crossover point are copied; two-point crossover, where all genes between two points are copied; and the more general k-point crossover which is performed over k different crossover points. In the most general case, every gene is independently crossed over with a specified recombination probability. 

Evolutionary approaches provide a simple, highly-parallelizable population-based approach to optimization. Since it only requires some measure of performance as a fitness function, the same algorithm can be effectively applied in contexts analogous to supervised, unsupervised or reinforcement learning. Additionally, the fact that these are population based methods enables better exploration of the parameter space in comparison to gradient-based methods and hence have a better shot at escaping local optima. Finally, evolutionary algorithms are only a part of a wide-range of nature inspired population-based stochastic search approaches optimization such as ant-colony optimization~\citep{dorigo1999ant}, particle-swarm optimization~\citep{kennedy1995particle}, and several others~\citep{beheshti2013review}. 

In an effort to benefit from the favorable features of both above-mentioned approaches, hybrid approaches of optimization have been developed. One such category of hybrid-approaches are population based gradient-estimation algorithms where performance in a cluster of parameter choices are compared to estimate the direction of the gradient without actually estimating the derivatives. These estimates have been done using as few as two data points, like in hill-climbing~\citep{russell2002artificial}; a population of randomly sampled data points, like in evolutionary strategies~\citep{beyer2002evolution}; or when possible, by considering all possible neighboring points, like steepest ascent hill-climbing~\citep{russell2002artificial}. Additionally, hybrid approaches are also employed in optimization methods that involve multiple time-scales. For instance, model parameters may be optimizing using gradient-based approaches, while model hyperparameters may be optimized using evolutionary approaches at a slower time-scale~\citep{young2015optimizing, loshchilov2016cma}. Finally, hybrid approaches can involve training different components of a model at different time-scales, which is especially suited when the individual components are mutually-dependent in their learning~\citep{ackley1991interactions, leite2020reinforcement}. Thus, hybrid approaches to learning span optimization at several levels and provide more sophisticated control over the process and have shown to be empirically beneficial compared to any one approach alone~\citep{todd2020interaction}. 

Altogether, several approaches to model optimization exist and of them, while no single approach is the best~\citep{wolpert1997no}, some approaches might better suit certain problems. For instance, search based approaches are better suited for discrete parameter spaces since gradients cannot be computed. Similarly, gradient-based approaches may lead to significantly faster convergence in convex landscapes. Alternatively, in work involving optimizing neural network models in a way that can explore all possible ways to solve a given problem it is crucial to adopt an approach that minimized the bias that an experimenter can introduce which naturally lends itself to evolutionary algorithms. In any case, a good approach to model optimization involves devising objective functions with known bounds, sufficient random exploration of the parameter space, and conducting several optimization runs to evaluate consistency and variance in results. 

\section{Information theory in Neuroscience}
Building neural network models and optimizing them to perform tasks gives us models of natural systems. While the resources required to achieve desired task performance is already informative of the natural system being modeled, these models are most useful when they are analyzed to understand them better. Approaches to understanding neural network operation in the context of behavior requires that it captures linear as well as non-linear interactions between the different components: environment-neural network and neuron-neuron. Several methodologies have been proposed to capture specific features of neural activity such as connectivity, effective dimensionality, encoding, and decoding. In this regard, information theory has emerged as a general framework to quantify stochastic properties and relationships between different variables in a system of interest. It provides tools to measure these quantities in a way that is invariant to the scale of the system and allows comparison across systems. In this section, I describe the main tools used in our analysis as well provide examples of other work within Neuroscience that had utilized information theoretic tools to study the neural basis of adaptive behavior. 

Information theory was first introduced by Claude Shannon in his seminal paper ``A mathematical theory of communication''~\citep{shannon1948mathematical}, as a methodology to develop efficient coding and communication of data across noisy channels.
Its rise to popularity can be primarily attributed to its ability to be applied in any domain, ranging from economics to Neuroscience.
It provides a set of tools that enable us to understand the relationships and interactions between arbitrary multivariate random variables. 
These tools enables us to answer questions such as, ``how can we quantify the difference in uncertainty in random process versus another?'', and ``how much does knowing the value of one random variable, reduce uncertainty about another?''. 
Since Shanon's introduction of information theory, there have been several advancements that allows study of information transfer over time and through the interaction of multiple sources of information. These advancements enable us to ask more involved questions such as, ``what is the amount of information transferred from one random process to another over time?'', and ``Of these multiple sources, what is the amount of information that is redundantly transferred from all sources about a target random variable?''. These crucial questions enable us to understand the interactions between different components of a complex system, and can ultimately lead to a mechanistic understanding of its operation.

A brief introductory primer to the most widely-used information theoretic methods is present in Appendix B. These information theoretic measures have been used extensively in Neuroscience in several contexts: \textit{in vitro}, \textit{in vivo}, and \textit{in silico} (or computational models). Neural responses are stochastic because of noise and because biological neural networks and their outputs are not purely a function of the input alone, but depend on their internal state as well~\citep{mainen1995reliability}. They could produce outputs in the absence of any input, or not produce any response to certain stimuli. This stochastic nature of their behavior makes information theory especially suited for analyzing and interpreting neural activity~\citep{dimitrov2011information, sayood2018information, jung2014applying, james2011anatomy, wibral2015bits}. 

\subsection{Estimating informational quantities}
One of the major challenges in utilizing information theoretic measures in experimental settings is the availability of sufficient data to infer the data distributions correctly~\citep{paninski2003estimation}. This is a complex problem and several parametric and non-parametric approaches have been proposed~\citep{silverman2018density}. To estimate data distribution from limited data, we have employed average shifted histograms for its beneficial trade-off between statistical and computational efficiency~\citep{scott1985averaged}. This involves discretizing the data space into a number of bins and estimating frequentist probabilities based on the bins occupied by data samples. To reduce the impact of arbitrarily chosen bin boundaries the data distribution is estimated by averaging the bin occupancies across multiple shifted binnings of the data space. This binning based estimator has been shown to approximate a triangle kernel estimator~\citep{scott1985averaged}. While the binning provides significant computational advantages, its approximation errors must be considered. Bias properties and guidelines for choosing the parameters for average shifted histograms are given in \cite{scott1985, fernando2009selection, scott1979optimal, scott2012multivariate}. For a moderate sample size, 5 to 10 shifted histograms has been shown to be adequate~\citep{scott1985}. In general, average shifted histograms are best suited for noisy continuous data where the distribution of the data is unknown. For a more involved discussion on density estimation and its bias properties we point the reader to~\cite{scott1987biased} and \cite{wand1994kernel}.

\subsection{Information and time}
One variable that is implicit in all formulations of information theoretic measures described above is time. Typically, these information quantities are measured disregarding the time variable. In other words, data distributions are estimated using data across all time-points. Consequently, the information measures are aggregate measures across the time duration over which data was collected. Alternatively, if data was collected in several trials these quantities can be measured \textit{in time}. Information thus measured as a function of time then reveals the dynamics of information in the system under study. This requires that there are a sufficient number of trials to develop reliable estimates of data distributions at each time point. This approach has been applied in several modeling studies as well as in experimental conditions. For instance, in our own previous work we measure predictive information in a neural network about a future environmental stimulus \textit{in time}~\citep{candadai2019sources}. The temporal analysis, unlike the time-averaged analysis, disambiguates the source of predictive information as being either the agent or the environment. We further demonstrate how information generated in the environment quickly gets reflected in the neural network in a way that cannot be distinguished using time-averaged analysis.

\subsection{Levels of Information theoretic analyses of neural networks}
In this section I present examples of work at different levels in which information theory has been applied starting from information processing in single neurons to information processing in neural networks engaged in closed-loop behavior. 

\subsubsection{Single Neurons}
A neuron is typically considered to be the smallest computational element in the animal brain. Although, this is in no way a statement trivializing the computational capacity of the neuron. As revealed by the landmark study of squids' giant axon, and its model~\citep{hodgkin1952quantitative}; as well as based on the ever increasing literature on dendritic computation~\citep{london2005dendritic}, neurons are rather sophisticated information processing systems. Characterizing tuning curves (stimulus-response relationship) has shown that neurons in the visual cortex are selective to orientation of stimulus~\citep{hubel1965receptive}, place cells in the hippocampus are selective to physical locations in space~\citep{o1978hippocampus}, cercal neurons in crickets are sensitive to wind direction~\citep{theunissen1991representation} and so on. Tuning curves were interpreted as neurons being most selective to stimuli that produced to maximum firing rate. However,~\cite{butts2006tuning} applied information-theoretic measures to demonstrate the validity of an alternate interpretation - the neuron is most selective to stimuli where the tuning curve has the highest slope. Measuring the specific information that noisy neural activity in medial temporal cortex had about specific values of the stimulus showed that information encoded was maximum near the peak-slope in the tuning curve. Intuitively, this is because there is greatest difference in firing rate for small changes to stimuli at this region of the tuning curve. Further, they showed how the experimental design (range and the number of stimuli provided) can change whether the maximum information lies at the peak firing rate or the highest slope. 

In addition to identifying the stimulus selectivity of a neuron, information-theoretic analyses have also been extensively applied by viewing neurons as communication channels, and hence estimating their information capacity or channel capacity. This is a measure of the number of distinctly identifiable stimuli that a neuron can encode, or mutual information between the input distribution and output distribution. Automatically, this requires that an assumption is made about the nature of encoding in order to build the output distribution from spiking activity. Two approaches to encoding that are widely accepted are rate coding and temporal coding~\citep{rieke1999spikes}. Once a specific encoding method has been chosen, the channel capacity will then depend on the noise in the channel, and the distribution of inputs using which the channel capacity will be estimated. Thus, this is a complex problem and even theoretical estimates will strongly depend on the assumptions made. Initial theoretical estimates made using the temporal coding scheme and a Gaussian noise distribution to the inter-spike intervals were made to be 4000 bits/s~\citep{mackay1952limiting, rapoport1960theoretical}. This was very optimistic estimate~\citep{stein1967information} was then re-evaluated using a more biologically realistic gamma distribution for the inter-spike intervals to demonstrate that the capacity was 15-50 bits/s with both the temporal and rate coding schemes~\citep{borst1999information, ikeda2009capacity}. These results put to rest on which coding scheme was most informative, and the debate now continues in the form of a energy-efficiency (temporal code) versus robustness (rate code). Like most phenomena in nature, animal brains perhaps use both under different conditions. 

\subsubsection{Populations of Neurons or Brain Regions}
Expanding the analysis from individual neurons to groups of neurons, information theory has been applied on two fronts: first, information encoded about stimulus can be measured in a population of neurons similar to how it is measured in a single neuron; and second, by measuring the information that is transferred from one neuron to another. 

Several advancements have been made in population coding and decoding approaches that show information being distributed in a population of neurons, and overlapping sensitive ranges across neurons leading to theories of robustness in encoding via population coding. For instance, information about the directionality of motor movement was shown to be best represented as a weighted average of several individual neuron responses~\citep{georgopoulos1982relations, georgopoulos1986neuronal, wessberg2000real}. Similarly, saccadic eye movements in monkeys were also shown to be controlled by the weighted sum of the activations of collicular neurons~\citep{lee1988population,sparks1976size}. While these studies were based on looking at peak-response curve values, an alternative information theoretic approach similar to the one described in the analysis of response curves in single neurons can be adopted. \cite{ince2010information} measured mutual information between the whisker stimuli and the population activity in rat somatosensory cortex for different population sizes. Their results showed that mutual information in populations of size 2 or greater captured the stimulus information very well, without any significant differences in larger populations. This led them to conclude that studying pair-wise interactions in neurons would be sufficient to characterize information transmission in neural networks. 

The idea that neural information processing can be understood by studying pair-wise interactions in a population is fundamental to the adoption of network theory in Neuroscience. This involves building interaction networks between neurons in population, as well as between brain regions, based on measuring the interactions between them pair-wise. This approach has flourished in the last decade under the new sub-field of \textit{Network Neuroscience}~\citep{sporns2005human, bassett2017network, sporns2014contributions}. While the literature on how to estimate these interaction networks is continuously growing~\citep{okatan2005analyzing,hlavavckova2007causality,pillow2008spatio,gerhard2011extraction}, Transfer entropy (TE) has emerged as one of the most widely used methods in Neuroscience because of its ability to infer directed edges between nodes in a network from node activity alone~\citep{wibral2014transfer}. As mentioned earlier, these nodes could be anything from individual neurons or macro brain regions. When estimated across all possible pair-wise node combinations, the \textit{effective network}~\citep{friston1994functional} is constructed. The effective network is representative of the interaction between neurons that results in the dynamics on the underlying structurally connected network.   

Studying properties of the effective network has yielded several insights on neural information processing, especially at the micro-scale. Multi-electrode array recordings of neural activity of organotypic cortical cultures \textit{in vitro} has shown the presence of rich-club architectures; some highly-connected neurons are more connected to each other than by chance~\citep{Shimono2015, Nigam2016}. This study further showed that information transfer in a neural network is non-uniform: 70\% of the information was propagated through only 20\% of the neurons. Following that, \cite{timme2016high} measured the synergistic information in a neuron's activity from the combined sources it receives information from, as an indicator of how much computation was performed by the neuron. Intuitively, this is a measure of how much more information is available in the neuron beyond just the union of information provided by the inputs. This analysis showed that computation in a neuron was more strongly correlated to the number of outgoing connections rather than the inputs it received. Furthermore, it was then shown by cross-referencing effective network properties with synergistic information that it was, in fact, the rich-club neurons that performed a disproportionate amount of computation, about 160\% more~\citep{faber2019computation}. 

In larger-scale neural dynamics, where ensemble neural dynamics corresponding to brain regions comprised of several thousand neurons are measured using fMRI, transfer entropy based effective networks can be constructed with nodes denoting brain regions. Such studies have yielded several insights as well into information transfer and its dynamics across the brain. For instance, \cite{Lizier:2011} estimated effective networks while subject performed a visuo-motor tracking task to show that TE was able to infer the expected directed connectivity between known motor planning regions in the brain. They also showed that increasing task-difficulty resulted in increased amounts of information transfer between the motor cortex and the cerebellum where fine-tuning of motor control occurs indicating the greater error correction happening as the task is more unpredictable. Similarly, by applying TE to magnetoencephalographic data collected during an auditory short-term memory experiment, \cite{Wibral:2011} showed that TE was able to capture the expected changes in information transmission left temporal pole and the cerebellum.

Altogether, the integration of network theory with information-theoretic measures is enabling the interpretation of neural activity to go beyond response curves in single neurons to understanding collective information processing in populations of neurons. 

\subsubsection{Neural Networks engaged in closed-loop behavior}
The neural network analyses described in the two previous sections are mostly using resting-state data where the neural network was exhibiting its intrinsic dynamics, or when a subject was passively performing a task. In the real-world animals are in closed-loop interaction with their environments where they act on the environment thereby influencing what they perceive. Perhaps the best examples of information theoretic analysis of neural networks that are involved in closed-loop behavior come from computational modelling studies. Information-theoretic analysis of embodied neural network models that were optimized to perform cognitively interesting tasks such as object categorization, relational categorization etc. have demonstrated that studying information flow through the integrated brain-body-environment system will provide a comprehensive understanding of the neural basis of behaviors~\citep{williams2008embodied, williams2010information, beer2015information, izquierdo2015information}. Such analysis also provide additional insights into how behavior is not entirely controlled by the brain because this allows for considering the environmental and body variables into the information theoretical analysis. For instance, using a model of embodied relational categorization \cite{williams2008embodied} showed that the optimized agents offloaded information storage to the environment. Agent's were optimized to behave differently in response to the relational category of ``smaller'' or ``larger'' based on a visual stimuli of objects falling in sequence. Amongst agents that solved this problem with similar levels of accuracy, some adopted a strategy of moving away from the falling object far enough such that their position was indicative of the size of the object. This behavioral characteristic was made mechanistic explanation by demonstrating the near-perfect amounts of information in the agent's position about the size of the perceived object. In a more biologically grounded model of \textit{C. elegans} klinotaxis, \citep{izquierdo2015} showed that although the underlying neural network parameters might be very different in different model instantiations, they were all identical in the pattern of information flow through the neural network. This was demonstrated by measuring time-varying mutual information in the neural activity about the behaviorally relevant variable, namely change in salt concentration to show that the ensemble of models that were analyzed all had similar information flow patterns. Similar work using time-averaged transfer entropy showed that the same structural network without any changes in parameters manifests itself as different functional networks when placed under different contexts thereby enabling performing multiple tasks~\citep{vasu2017evolution}. Thus, such an analysis allows us to go beyond the individual structural differences, and study the common functional characteristics of neural networks thereby enabling us to interpret neural activity in the context of behavior. Altogether, information theoretic analyses of brain-body-environment models takes us one step closer to a comprehensive understanding of the neural basis of adaptive behavior.

\subsubsection{Multi-agent models}
Taking the embodied cognition perspective one further step involves incorporating reactive elements in the environment (namely, other independent agents). Now a single agent with $N$ neurons is viewed simply as a slice of a higher-dimensional system with $MxN$ neurons where $M$ is the number of agents. Information theoretic analysis of multi-agent systems has provided insights into the relationships between neural and behavioral complexity and how active social interaction plays a crucial role in enabling complex behaviors~\cite{candadai2019embodied, resendiz2020enhanced, resendiz2020levels}. While, this is a relatively less explored area, there is work that uses information theory to study social factors of decision making~\citep{frith2008role}, intersubjective experience~\citep{aston2019metaplasticity} and so on.

\subsection{Tools for information theoretic analyses}
Several researchers have developed software packages that aid information theoretic analyses. Pyentropy is a python package that allows easy estimation of entropies~\citep{inceinformation}. JDIT is a JAVA package that was primarily designed for measuring transfer entropy but also includes other measures such as entropies and mutual information~\citep{lizier2014jidt}. TRENTOOL is a popular MATLAB toolbox which primarily caters to analog neural data such as MEG~\citep{Lindner:2011}. \cite{Ito:2011} presented a MATLAB toolbox for TE estimation from binary spiking data that accounted for several delays and picked the most reliable transfer entropy value using a data-shuffled baseline. \textit{infotheory} is a packaged we developed for PID measures as well as transfer entropy estimation~\citep{candadai2019infotheory}. Two existing packages that are most similar to ours are dit~\citep{james2018b} and IDTxL~\citep{wollstadt12019}. Unlike dit, our package can also help analyze continuous-valued data and unlike dit and IDTxL we have implemented PID analysis of 4 variables: 3 sources and 1 target. In light of these existing packages and their functionalities, our package primarily focuses on measuring multivariate informational quantities on continuous data where the data distribution is not known a priori. However, it can still be used with discrete data using the same methods.

\section{Conclusion}
This work discusses the use of computational models and information theory to advance our understanding of the neural basis of behaviors. Computational models provide the ability to study our biological systems of interest at a level of abstraction that enables analyses in a way that is not tractable in biological systems. They provide complete access to all variables of a system and, more so, the ability to manipulate them in a way that experimentalists cannot. Information theory, on the other hand, provides the tools for an in-depth analysis of these models.
Insights obtained from analysis of the model advance our theoretical understanding of neural information processing. It provides proof-of-existence, and general principles underlying neural information processing in adaptive behavior. Additionally, information-theoretic tools can acts as the bridge between modeling and experimentation. Analyses performed on an idealized model can be directly applied to appropriate experimental data; with the confidence that model-based evaluation of the methods have demonstrated the efficacy of the method to capture the phenomena of interest. 
Ultimately, the combination of computational modeling and information theory is a potent combination that can advance theoretical as well as experimental Neuroscience for years to come. 

Our primary research goal is to conduct information-theoretic analyses of computational models of neural networks to advance our understanding of the neural basis of behavior. Consequently, it becomes crucial to identify the level of abstraction at which we build models to study the neural basis of adaptive behaviors. Several perspectives exist in this regard, and perhaps the most widely adopted is from mainstream theoretical Neuroscience where models are built primarily at one of two levels: sensory-response models to describe what the input-output relationship is between stimuli and neural activity; and circuit-level models that describe how these relationships maybe implemented using known physiological data~\citep{abbott2008theoretical}. However, these approaches limit cognition to only the brain. There is an increasing realization that the role played by the environment and the body in behavior needs to be take seriously in order to develop a comprehensive understanding of the neural basis of behavior~\citep{krakauer2017neuroscience, chiel1997brain, Varela1991}. This is inline with approaches in Computational Neuroethology where models are built not only of neural circuitry, but also of the body, and ecological context of the model system, all of which are considered to be equally important in behavior~\citep{datta2019computational, chiel2008computational, Pfeifer2007}. There have been two prominent approaches to building models from a neuroethological perspective: bio-robotics and evolutionary robotics. Bio-robotics, as the name suggests, involves building robots as models of specific animal systems to test and generate hypotheses regarding the control of behavior~\citep{webb2002robots, webb2001can, beer1993biological}. Building robots ensures that the physical aspects of an animal's ecological context are faithfully reproduced. Further, robots that are models of specific animals enable targeted understanding of behavior in that animal. For example,~\cite{moller1998modeling} built a robotic model of desert ant navigation to demonstrate that it was a combination of path-integration and visual piloting that enabled these ants to navigate to precise locations over large distances. Similarly,~\cite{ijspeert2007swimming} built a robotic salamander to demonstrate that the same neural oscillators when modulated by environmental feedback can produce the walking and swimming gaits observed in salamanders. These studies not only act as existence-proofs for mechanisms that drive specific behaviors in a target mechanism, but also provide specific testable hypotheses for the animals that the robots are based on. The evolutionary robotics approach, on the other hand, involves building simulated idealistic models of artificial agents with emphasis on building minimal models of cognition that enable in-depth analysis~\citep{harvey2005evolutionary, cliff1993explorations, beer1992evolving}. Integral to evolutionary robotics is use of evolutionary algorithms to optimize model parameters, an approach that minimizes experimenter bias by optimizing neural networks based on \textit{what} they are optimized to do and not \textit{how} the experimenter thinks it should be done. Crucially, in contrast to bio-robotics models, these models are typically not built to physiological data from a specific animal. This is done because freeing the constraint of adherence to a specific model organism enables the construction of minimal artificial agents that nevertheless performing cognitively interesting tasks are tractable for in-depth analysis, and exploring general principles in behavior that go beyond generating hypotheses for specific animals~\citep{beer2009animals}. For instance, using a minimal model of a 1-dimensional agent controlled by a dynamical neural network, \cite{beer2015information} show that embedded, embodied agents perform a task that requires memory without an explicit internal storage mechanism; the agent offloaded memory to its position in the environment: a testament to the benefits of adopting a neuroethology perspective. Similarly,~\cite{izquierdo2008analysis} showed that dynamical neural networks can perform multiple tasks even in the absence of neuromodulation or plasticity. Overall, building models that incorporate the environment and agent-environment interaction expand the range of mechanisms that can be discovered to produce behaviors, and utilizing evolutionary algorithms further aid this by enabling the generation of integrated sensorimotor systems with minimal bias. Computational models are especially suited when this approach is taken to research because, although there have been developments in tools for experimental data acquisition, there are still challenges in reliably recording all required data from a freely moving animal. In such cases, theoretical advancements must not wait for experimental technologies to arrive but can instead precede them by advancing our understanding of general principles of adaptive behavior through the analysis of computational models. 

In summary, the research methodology presented in this paper takes the approach of developing computational models of dynamical recurrent neural networks, optimizing them using an evolutionary algorithm to perform tasks that involve embedding or embodying the neural network models, and followed by information-theoretic analysis of the optimized models. In all cases, it is preferable that an ensemble of models are built and analyzed to check if the results are consistent or if several solutions exist that solve the same problem. While this is a specific instantiation of the type of computational models and the tool for analysis, it is by no means constraining. Ultimately, we believe that insights from analyzing these models can serve as hypotheses that guide experimental research.

\bibliographystyle{unsrtnat}
\bibliography{template}  






\clearpage
\setcounter{section}{0}
\renewcommand*{\theHsection}{chX.\the\value{section}}
\section{Appendix A: Neuron and neural network models}
Computational models of neurons and neural networks have been built at many different levels of abstraction, on a spectrum from biophysically realistic models to simplified linear models. These models are also on a complementary spectrum of computational efficiency spectrum where the most biophysically realistic models are the most computationally expensive to linear models that are least computationally expensive. Several models exist on this spectrum and models can be appropriately chosen that is a trade-off between computational efficiency and biophysical realism. Some of the key characteristics that decide where a model would lie on these spectra are: spiking versus non-spiking models, with or without internal state, and recurrent versus feed-forward connectivity. Within neuron models that have an internal state, there is a distinction between continuous- versus discrete-time systems that are defined by differential and difference equations respectively. While this is a theoretical distinction, the distinction is not significant in simulation since continuous-time systems are also simulated in digital computers using discrete approximations such as Euler integration. In this section, I describe the widely-used neuron models and explain where they lie on the spectrum of biophysical realism as well as computational efficiency in relation to other models.

At the very end of the biophysically realistic spectrum lies the Hodgkin-Huxley model~\cite{hodgkin1952quantitative}. It was published in 1952 by Alan Hodgkin and Andrew Huxley based on perhaps the most extensive studies of biological neurons, specifically the squid's giant axon. A single biological neuron is modeled as an electrical circuit where parameters of ion flow across the cell membrane is modeled using electrical analogues such as resistors and capacitors. This non-linear continuous-time model exhibits spiking activity extremely identical to to biological neurons when injected with external current. As such, it is a spiking neuron model defined by continuous-time differential equations, and is on the far end of the biophysically realistic models side of the spectrum. Consequently, simulating a single neuron per this model is quite expensive computationally. Since the model is that of a single neuron, it does not place an emphasis on how neurons are connected to one another and therefore theoretically allows any kind of connectivity. Following the Hodgkin-Huxley model, a simplified version of it was proposed in 1961 by Richard FitzHugh~\citep{fitzhugh1961impulses}, and later augmented with a circuit model by~\cite{nagumo1962active}. With the main motivation of developing a model of the injection and propagation of current through a neuron, this simplified version abstracted away from independently modeling individual ion flows. For weak external excitation, the model produces activity that resembles subthreshold membrane potential dynamics. With a sufficiently high external current, the model produces spikes. Importantly, the model does not exhibit all-or-none spiking activity: for intermediate external excitation, the model produces spikes of smaller magnitudes. Finally, with consistently high external stimulation, the model produces oscillatory spiking activity. With only 2 state-variables as opposed to Hodgkin-Huxley's 4, this model enables a visualization of the entire phase-space of the neural dynamics and its phase-space has been analyzed extensively~\citep{izhikevich2008dynamical}. Since the model has abstracted away from modeling ion-channels in detail, it is computationally more tractable than the Hodgkin-Huxley model. It is still biophysically realistic in its ability to produce spiking and refractory dynamics similar to biological neurons. While these two models form the foundation for biophysical modes of neuraons, several others have been proposed at different levels of simplification: Izhikevich spiking model~\cite{Izhikevich2003}, Integrate-and-fire family of models~\citep{abbott1999lapicque}, Continuous-Time Recurrent Neural Network (CTRNN) Models ~\citep{Beer:1995}, Long-Short Term Memory discrete time models~\citep{hochreiter1997long}, and the discrete time state-less Perceptron model~\citep{rosenblatt1957perceptron}. 

From this non-exhaustive list of neuron models at different levels of abstraction and computational complexity, appropriate models can be selected depending on the purpose for which the model is built. For instance, modeling specific brain regions such as hippocampal neurons requires that a biophysically realistic model of spiking neurons is chosen with appropriate computational complexity: integrate-and-fire neurons. Conversely, if the goal of building the models is to solve an engineering problem, biophysical realism might take a backseat giving way to LSTMs or multi-layer perceptrons as the model of choice. Finally, understanding general principles underlying neural information processing in living organisms does not necessarily have to use spiking neuron models and can employs CTRNNs as approximations of rate-coding. Of these, the Izhikevich and CTRNN models lie at an optimal trade-off between biophysical realism and computational tractability. These models are discussed in detail below. 

\subsection{Izhikevich model}
As a follow-up to the two-dimensional FitzHugh-Nagumo model, Eugene M. Izhikevich published his model in 2003~\citep{Izhikevich2003}. This model of spiking neurons included and explicit threshold based spiking in order to produce all-or-none spiking dynamics unlike that of the FitzHugh-Nagumo model. Mathematically, this model defines rates of change in membrane potential, $V$, and a recovery variable, $U$, according to

\begin{equation}
    \begin{split}
        \Dot{V} &= 0.04V^2 + 5V + 140 - U + I\\
        \Dot{U} &= a(bV - U)
    \end{split}
\end{equation}

\noindent with explicit threshold based spiking and reset of state-variables as follows 

\begin{equation}
    \text{if v }\geq \text{30 mV,\ then}
    \begin{dcases}
        v \leftarrow c \\
        u \leftarrow u + d
    \end{dcases}
\end{equation}

\noindent where $a$, $b$, $c$, and $d$ are parameters whose ranges have been thoroughly studied and mapped to characteristic behaviors. As such, this model provides an optimal trade-off between computational complexity and biophysical realism, because it is a 2-D continuous-time, spiking neuron model that can replicate various spiking dynamics observed in cortical neurons such as intrinsic bursting, low-threshold spiking and so on. 

This model's utility has been demonstrated in several ways: its ability to perform a range of linear and non-linear pattern recognition tasks~\citep{vazquez2010izhikevich}, as models of specific brain regions~\citep{prescott2006robot}, as generic models of cortical activity to validate analytical methods~\citep{ito2011extending}, as well as neural controllers in brain-body-environment models of adaptive behavior~\citep{vasu2017information}.

\subsection{Continuous-Time Recurrent Neural Network model}
Moving from spiking to non-spiking neuron models, Continuous-Time Recurrent Neural Networks (CTRNNs) are models of dynamical recurrent neural networks with internal state. CTRNNs are perhaps the simplest non-linear, continuous dynamical neural network models, and yet are universal dynamics approximators~\citep{funahashi1993approximation}. As their name denotes, they are continuous-time models and are recurrently connected. Thus, they are an obvious choice to build computational models of neural networks to perform tasks that involve stateful interaction with the environment like living organisms do.  and a network is defined a vector differential equation:

\begin{equation}
    \tau \Dot{y} = -y + W\sigma(y + \theta) + I
\end{equation}

\noindent where $\tau$, the time-constants, $y$, the internal-states of the neurons, $\theta$, the biases of the neurons and $I$, the external input are all $N$ dimensional vectors corresponding to the $N$ neurons that make up the circuit. $W$ is a synaptic weights matrix denoting the fully-connectedness of the circuit, and $\sigma$ is the sigmoidal function whose value gives the output of the neurons. 
If interpreted as a spike-rate model, $\Dot{y}$ denotes the membrane potentials, $\sigma(.)$, the mean firing rates, $\theta$, the firing thresholds, and $W_{ii}$, the self-weights represents a simple active conductance. 

The fact that that they are not spiking models does not entirely discount their place on biophysical realism because there has been evidence of non-spiking neurons in locusts~\citep{siegler1979morphology}, crayfish~\citep{takahata1981physiological}, and most neurons in \textit{C. elegans}~\citep{lockery2009quest}. If interpreted as models of non-spiking biological neurons, $\sigma(.)$ can instead represent the saturating non-linearities in the synaptic inputs.

This model's utility has been demonstrated in several ways: to approximate a variety of dynamical systems~\citep{funahashi1993approximation}, to build biologically-grounded models of living systems~\citep{izquierdo2019role}, as generic models of neural dynamics to validate analytical methods~\citep{candadai2019sources}, as well as neural controllers in brain-body-environment models of adaptive behavior~\citep{beer1992evolving}.

\subsection{Hodgkin–Huxley model}
The Hodgkin-Huxley model~\cite{hodgkin1952quantitative} was published in 1952 by Alan Hodgkin and Andrew Huxley. It based on perhaps the most extensive studies of biological neurons, specifically the squid's giant axon. A single biological neuron is modeled as an electrical circuit where parameters of ion flow across the cell membrane is modeled using electrical analogues such as resistors and capacitors. The mathematical description of a single neuron is as follows: the current flowing through the cell membrane, $I_c$, is modeled as a capacitance across the membrane, $C_m$, with potential difference across the membrane denoted by $V_m$, 

\begin{equation}
    I_c = C_m\frac{dV_m}{dt}
\end{equation}

\noindent in addition to $I_c$, the current flowing through an ion-channel, $I_i$, is modeled based on a resistor, or more specifically, an inverse-resistor or conductance, $g_m$ as follows

\begin{equation}
    I_i = g_m (V_m - V_i)
\end{equation}

\noindent where $V_i$ is the reverse potential of the ion channel (Sodium or Potassium). Thus, the total current through the membrane is given by the sum of the current across the lipid bilayer, Sodium ion currents, Potassium ion currents, and finally another ionic current term to account for leak currents, as follows:

\begin{equation}
    I = C_m\frac{dV_m}{dt} + g_{Na} (V_m - V_{Na}) + g_K (V_m - V_K) + g_l (V_m - V_l) 
\end{equation}

\noindent where $g_l$ and $V_l$ correspond to the conductance and voltage for the leak currents. 

In addition to modeling current flow across the membrane, the model defines the activations of ionic channels also as differential equations

\begin{equation}
    \frac{dn}{dt} = \alpha_n(V_m) (1-n) - \beta_n(V_m) n
\end{equation}

\begin{equation}
    \frac{dm}{dt} = \alpha_m(V_m) (1-m) - \beta_m(V_m) m
\end{equation}

\begin{equation}
    \frac{dh}{dt} = \alpha_h(V_m) (1-h) - \beta_h(V_m) h
\end{equation}

\noindent where $n$, $m$, and $h$ describe the dynamics of potassium channel activation, sodium channel activation and sodium channel inactivation respectively; $\alpha_i$ and $\beta_i$ are rate-constants that are functions of $V_m$ given by 

\begin{equation}
    \begin{split}
         \alpha_n(V_m) &= \frac{0.01(10-V_m)}{\exp\big(\frac{10-V_m}{10}\big)-1}\\ 
         \alpha_m(V_m) &= \frac{0.1(25-V_m)}{\exp\big(\frac{25-V_m}{10}\big)-1}\\
        \alpha_h(V_m) &= 0.07\exp\bigg(\frac{-V_m}{20}\bigg)\\
        \beta_n(V_m) &= 0.125\exp\bigg(\frac{-V_m}{80}\bigg)\\
        \beta_m(V_m) &= 4\exp\bigg(\frac{-V_m}{18}\bigg)\\
        \beta_h(V_m) &= \frac{1}{\exp\big(\frac{30-V_m}{10}\big) + 1}
    \end{split}
\end{equation}

The Hodgkin-Huxley model has 4 state-variables $I$, $n$, $m$, and $h$ that define a single neuron's activity. This non-linear continuous-time model exhibits spiking activity extremely identical to to biological neurons when injected with external current. As such, it is a spiking neuron model defined by continuous-time differential equations, and is on the far end of the biophysically realistic models side of the spectrum. Consequently, simulating a single neuron per this model is quite expensive computationally. Since the model is that of a single neuron, it does not place an emphasis on how neurons are connected to one another and therefore theoretically allows any kind of connectivity. This model won them the nobel prize for contributions to Physiology or Medicine.

\subsection{FitzHugh-Nagumo model}
Following the Hodgkin-Huxley model, a simplified version of it was proposed in 1961 by Richard FitzHugh~\citep{fitzhugh1961impulses}, and later augmented with a circuit model by~\cite{nagumo1962active}. With the main motivation of developing a model of the injection and propagation of current through a neuron, this simplified version abstracted away from independently modeling individual ion flows and replaced it with state variables for the membrane voltage, $V$, and total current, $I$; and introduced a recovery variable, $W$. 

\begin{equation}
    \begin{split}
        \Dot{V} &= V - \frac{V^3}{3} - W + I\\
        \tau \Dot{W} &= V + a - bW
    \end{split}
\end{equation}

\noindent where $I$ refers to the externally injected current; and $a$, $b$ and $\tau$ are parameters that describe the rate of recovery of the membrane potential. 

This model was inspired by the van der Pol equations for a nonlinear relaxed oscillator. For weak external excitation, the model produces activity that resembles subthreshold membrane potential dynamics. With a sufficiently high external current, the model produces spikes. Importantly, the model does not exhibit all-or-none spiking activity: for intermediate external excitation, the model produces spikes of smaller magnitudes. Finally, with consistently high external stimulation, the model produces oscillatory spiking activity. With only 2 state-variables as opposed to Hodgkin-Huxley's 4, this model enables a visualization of the entire phase-space of the neural dynamics and its phase-space has been analyzed extensively~\citep{izhikevich2008dynamical}.

Since the model has abstracted away from modeling ion-channels in detail, it is computationally more tractable than the Hodgkin-Huxley model. It is still biophysically realistic in its ability to produce spiking and refractory dynamics similar to biological neurons. 


\subsection{Integrate-and-fire family of models}
Perhaps the most widely used spiking neuron model, the integrate-and-fire model is the computationally least expensive. Membrane potential, $u$'s dynamics is modeled using a single differential equation that is a combination of resistive and capacitive components connected in parallel to model the cell membrane, $u$.  

\begin{equation}
    \tau \frac{du}{dt} = -u + R I
\end{equation}

\noindent and all-or-none spiking activity is elicited based on a threshold as follows 

\begin{equation}
    \text{if u} \geq u_{th},\ u = u_{rest}
\end{equation}

\noindent where $\tau$ refers to the time-constant of the RC circuit, modeling the cell membrane with resistance, $R$; $I$ is the input current; $u_{th}$ is a threshold voltage for spiking, and $u_{rest}$ is the resting membrane potential. When the membrane potential, $u$, exceeds the threshold, a spike is registered, although the spiking activity is not an explicit part of the model. Following that, often an absolute refractory period is added during which the membrane potential stays at, $u_{rest}$.

Intuitively, this model of a spiking neuron is an intrinsically leaky dynamical system that integrates incoming current until the threshold is reached. Therefore, it is also often referred to as the leaky integrate-and-fire neuron model. This 1-dimensional dynamical system has been adopted widely because of its computational simplicity, however, it cannot produce certain types of spiking dynamics such as bursting. To this end, several modifications have been proposed to this model, a few key ones are described next. 

The integrate-and-fire-or-burst model~\citep{smith2000fourier}, that allows for bursting dynamics through the addition of a state-variable, $h$. \cite{ermentrout1996type} developed the quadratic integrate-and-fire model where the membrane potential decayed quadritcally with respect to its deviation from the rest and threshold voltages. This integration of $u_{rest}$ and $u_{th}$ in to the differential equation meant that the neurons could now exhibit activity-dependent thresholds, as well as exhibit a bistability of resting and tonic spiking modes. Finally,~\cite{mihalacs2009generalized} developed the generalized integrate-and-fire model with at least three state variables, the membrane voltage, the instantaneous threshold, and an arbitrary number of internal currents. This model was shown to produce a variety of spiking dynamics such as tonic bursting, phasic spiking and so on. Networks built of any of these neuron models do not have any constraints on how they are connected: recurrent or feed-forward. Overall, integrate-and-fire models lie on its own spectrum but are generally among the cheaper computationally at the cost of biophysically realistic spiking behavior. 

\subsection{Long-Short Term Memory model}
Entering the foray of discrete-time models, Long-Short Term Memory (LSTM) models are widely popular models in the artificial intelligence community because they have an internal-state, are recurrent, and hence able to integrate information across time-scales~\citep{hochreiter1997long}. The output, $h_t$, and internal state, $C$, of an LSTM cell evolves as follows:

\begin{equation}
    \begin{split}
        C_t = \sigma(W_f [h_{t-1}, x_t] &+ b_f) C_{t-1} + \\
        &\sigma(W_i [h_{t-1}, x_t] + b_i) * tanh(W_C [h_{t-1}, x_t] + b_C)
    \end{split}
\end{equation}

\begin{equation}
    h_t = \sigma(W_o [h_{t-1}, x_t] + b_o) * tanh(C_t)
\end{equation}

\noindent where $x_t$ is the input at time $t$, and all $W_i$ and $b_i$ refer to weights and biases corresponding to different ``gates'' within the cell: $W_f$ and $b_f$ form the forget gate, which regulates the extent to the which previous output of the cell is remembered going forward; $W_i$ and $b_i$ form the input gate that acts as a weighted mask to be applied over the candidate values generated by $W_C$ and $b_C$ to update the internal state; and finally $W_o$ and $b_o$ regulate the transformation of the internal state into the output of the cell. Note that all these weights are applied to the concatenated input and internal state vector. Thus, these gates are state- and context-dependent in how they modulate the updating of internal states and generation of outputs. 

Since they are defined by difference equations and not differential equations, they are theoretically less computationally complex. However, practically using an LSTM at a time-resolution similar to that of integrating a continuous-time system would make them more complex than CTRNNs due to the more sophisticated operations within an LSTM cell. LSTM gates do not have any direct neurobiological interpretation, however, they capture some of the requisite functional characteristics of biological neurons such as the presence of internal state, and recurrence. As such, these models have been used widely for tasks that require integrating information over long time windows. 

\subsection{Perceptron model}
Finally arriving to the other end of the spectrum, the perceptron model is the computationally lease expensive but the least biologically realistic. The model proposed by McCulloch and Pitts in 1943~\citep{mcculloch1943logical}, was generalized by Frank Rosenblatt in 1957~\citep{rosenblatt1957perceptron} as the Perceptron. The perceptron is a discrete-time model, with no internal state. It is a purely reactive model where the outputs, $o$, are a function only of the current inputs, $x$, and the network weights, $W$ and bias $b$ as follows:

\begin{equation}
   o = f(Wx + b) 
\end{equation}

\noindent where $f(.)$, the activation function can be any linear or non-linear function, thereby defining the family of functions that can be implemented by a perceptron. Examples of activation functions include threshold-functions, sigmoid, tanh or ReLU. 

Perceptrons can be stacked in multiple layers as feed-forward networks or they can be recurrently connected to induce an internal state through network activation. Multi-layer perceptrons have been extremely widely adopted due to their computational tractability in a variety of engineering problems forming the foundation for all deep neural networks.

\section{Appendix B: Information theoretic measures}
This section provides a quick introduction to the information theoretic tools. The relevance of each of these measure depends on the data, its source, and the motivation behind the analyses. Refer to~\cite{cover2012elements} for a more detailed account of these measures. More recent papers that introduce information theory to Neuroscientists are \cite{mcdonnell2011introductory} and \cite{timme2018tutorial}.

\subsection{Entropy}
Put simply, entropy can defined as a measure of uncertainty of a random variable. Greater the entropy, the more difficult it is to guess the value the random variable might take. It is function of the probability distribution rather than the actual values taken by the random variable, and it is estimated as follows - 

\begin{equation}
    H(X) = -\sum_{x \in X} p(x)\ log\ p(x)
\end{equation}

\noindent where $H(X)$ denotes entropy of the random variable, $X$, the summation is over all values the random variable can take, $\forall x \in X$ and $p(x)$ or $p(X=x)$ denotes the probability that the random variable $X$ takes a particular value $x$. The logarithm is base 2 and so the measured entropy is in the unit of bits. Note that $0log0$ is taken to be $0$, and therefore adding new terms to $X$ with probability $0$ does not change its entropy.

Intuitively, for a given random variable, a uniform probability distribution over its values would result in the highest entropy since every value is equally likely, making it most difficult to guess. On the other hand, other ``peaky'' distributions such as Gaussian would have lower entropy since most of the probability mass is near the mean. From a communication systems perspective, where information theory had its origins, entropy can also be interpreted as the average number of bits required to efficiently encode the random variable. For example, a fair coin requires 1 bit (0-heads, 1-tails); a uniform distribution over 8 values of a random variable requires 3 bits, but a non-uniform distribution could possibly be encoded by smaller average number of bits by assigning shorter codes for more likely outcomes. 

\begin{figure}[t]
    \centering
    \includegraphics[width=0.5\textwidth]{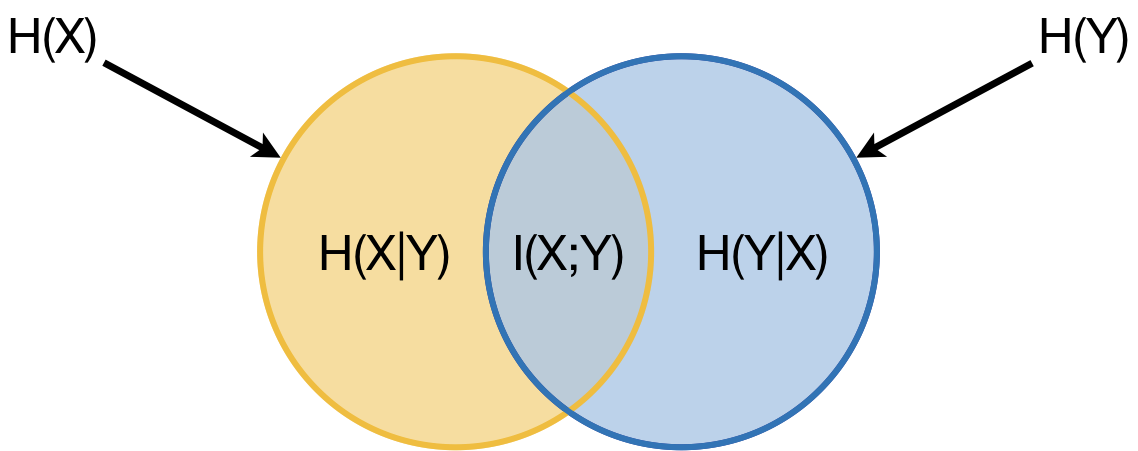}
    \caption{Relationship between entropy, conditional entropy and mutual information.}
    \label{fig:entropy_mi}
\end{figure}

\subsection{Mutual Information}
One of the most widely used information theoretic measures is mutual information. It is a measure of the amount of information two random variables have about one another - it is a symmetric measure. The information one variable has about the other can be expressed as the reduction in uncertainty about one variable upon knowing the other i.e. the difference between entropy of the variable and the entropy given the other variable. 

\begin{equation}
    \begin{split}
    I(X;Y) = I(Y;X) = &= H(X) - H(X|Y)\\
            &= H(Y) - H(Y|X)\\
    \end{split}
\end{equation}

\noindent where $I(X;Y)$ is the mutual information between random variables $X$ and $Y$, $H(X)$ and $H(Y)$ are their respective entropies, and $H(X|Y)$ and $H(Y|X)$ are their corresponding conditional entropies. The conditional entropy can be estimated from the conditional probability density, which is related to their joint probability density. Upon writing out the entropy expressions about in terms of the marginal and joint densities we arrive at the following expression for mutual information between $X$ and $Y$

\begin{equation}
    I(X;Y) = \sum_{x \in X} \sum_{y \in Y} p(x,y)\ log \frac{p(x,y)}{p(x)p(y)}
\end{equation}

The relationship between the individual entropies, the conditional entropies and mutual information can be better understood using a Venn diagram, as shown in figure~\ref{fig:entropy_mi}.

\subsection{Specific Information}
While mutual information measures the reduction in uncertainty in $X$, across all values of $Y$, specific information is a measure of reduction in uncertainty for any given value off $Y=y$. Specific information $I_{spec}$ is defined as follows~\cite{DeWeese:1999}

\begin{equation}
\label{eqn:Ispec}
    I_{spec}(X;Y=y) = \sum_{x \in X}{p(x|y)[log(\frac{1}{p(y)})-log(\frac{1}{p(y|x)})]}
\end{equation}

Based on this formulation, mutual information can be defined as a aggregate of specific information as follows

\begin{equation}
    I(X;Y) = \sum_y p(Y=y) I_{spec}(X;Y=y)
\end{equation}

\subsection{Partial Information Decomposition}
\begin{figure}[t]
    \centering
    \includegraphics[width=0.9\textwidth]{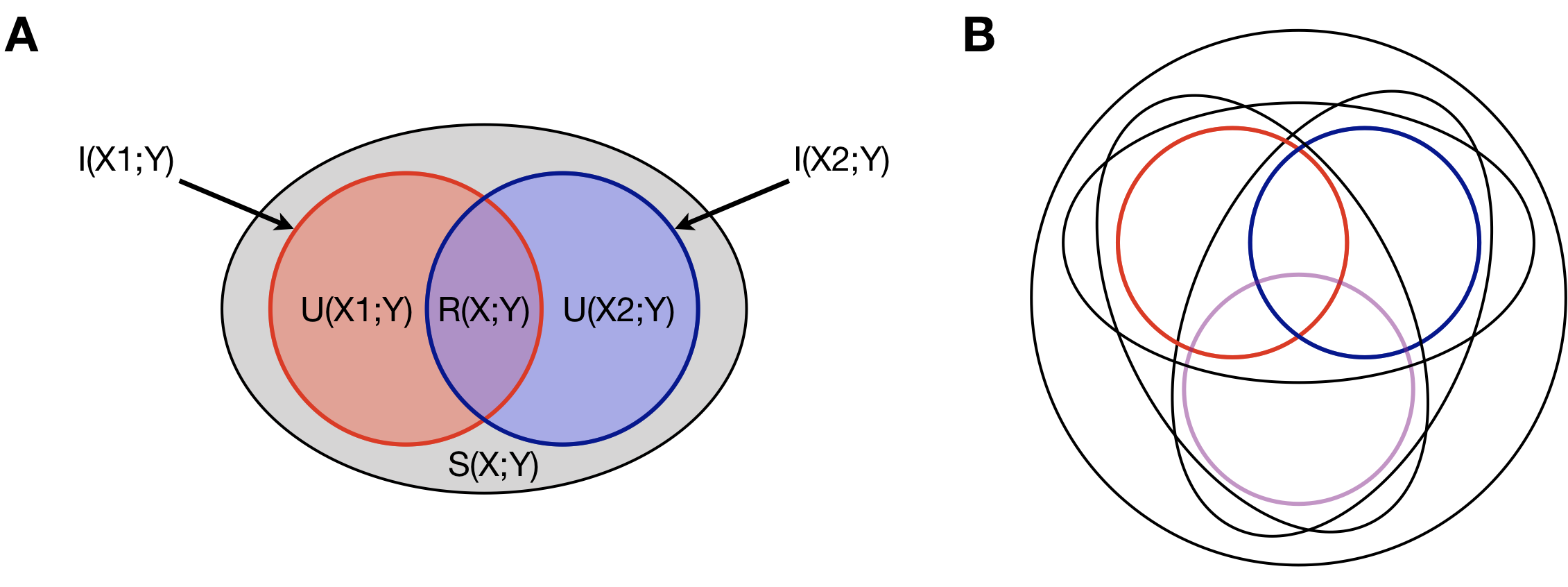}
    \caption[Non-negative partial information decomposition of total mutual information.]{Non-negative partial information decomposition of total mutual information. [A] In the case of 3 variables (2 sources $X=[X1, X2]$ and 1 target, $Y$), each source provides unique information about the target, $U(X1; Y)$ and $U(X2; Y)$, they provide information about the target redundantly, $R(X, Y)$ and finally they provide synergistic information that is only available from the combined knowledge of both sources, $S(X; Y)$.}
    \label{fig:pid}
\end{figure}

When there are multiple sources of information (or even a random variable that is 2-dimensional or more) about another random variable, the total mutual information between these sources and the `target' variable can be decomposed into its non-negative constituents, namely, unique information from each source, redundant information that the sources provide, and synergistic information due to combined information from multiple sources~\cite{williams2010nonnegative}. Consider two sources of information (two random variables) $X1$ and $X2$, about the random variable $Y$. If $X=\{X1,X2\}$, the total mutual information $I(X;Y)$ can be written as follows 

\begin{equation}
    I(X;Y) = U(X1;Y) + U(X2;Y) + R(X;Y) + S(X;Y)
\end{equation}

\noindent where $U$ denotes unique information, $R$ denotes redundant information and $S$ denotes synergistic information from these sources about $Y$. The decomposition can be better understood when visualized as a Venn diagram as shown in figure~\ref{fig:pid}. Naturally, with more than two sources, redundant and synergistic information will be available for the all combinations of the different sources. Also, note that each random variable used in these descriptions could be a multi-dimensional. The number of sources are just individual data from the system and can be of any dimensioniality.

\textbf{Redundant Information}
The sum of the minimum value of specific information each source provides is defined as the redundant information from the two sources. In other words, continuing the $(\{X1,X2\},Y)$ example, for each value of $Y$, the specific information $X1$ and $X2$ provide is independently computed, its minimum is found and this values is summed across all values of $Y$.

\begin{equation}
        R(X;Y) = \sum_{y\in Y} p(Y=y) min\{I_{spec}(X1;Y=y), I_{spec}(X2;Y=y)\}    
\end{equation}
 
where specific information is estimated as follows 

\begin{equation}
        I_{spec}(X1,Y=y) = \sum_{x\in X1} p(X1=x,Y=y) log\frac{p(X1=x,Y=y)}{p(X1=x)p(Y=y)}
\end{equation}

\textbf{Unique Information}
The amount of information that each source uniquely contributes, the unique information from that source, is estimated as the difference between the mutual information between that source and the `target' variable and the redundant information estimated as shown in the previous section. This is also apparent from the Venn diagram representation shown in figure~\ref{fig:pid}.

\begin{equation}
    \begin{split}
        &U(X1;Y) = I(X1;Y) - R(X;Y)\\
        &U(X2;Y) = I(X2;Y) - R(X;Y)
    \end{split}
\end{equation}

\textbf{Synergistic Information}
The information about $Y$ that does not come from any of the sources individually, but due to the combined knowledge of multiple sources, is synergistic information. For example, in an XOR operation, knowledge about any one of the inputs is insufficient to predict the output accurately, however, knowing both inputs will determine the output. In contrast, in an AND gate, merely knowing that one of the inputs is $False$ can lead to the conclusion that the output is $False$. Again, based on the Venn diagram in figure~\ref{fig:pid}, synergistic information from sources $X={X1,X2}$ about $Y$ can be determined based on total, unique and redundant information components as follows

\begin{equation}
    S(X;Y) = I(X;Y) - U(X1;Y) - U(X2;Y) - R(X;Y)
\end{equation}

\subsection{Transfer Entropy}
Given two random processes, $X$ and $Y$, transfer entropy from $X$ to $Y$, $TE_{X\rightarrow Y}$ is defined as the information transferred from from past values of $X$, $X_{t-d}$ to $Y_t$, over and above that information that $Y$'s own past, $Y_{t-k}$ provides. Intuitively, transfer entropy is a measure of thee reduction in uncertainty that know the source variable provides over the about the target variable's own reduction in uncertainty. 

\begin{equation}
    TE_{X\rightarrow Y} = I(Y_t; X_{t-d}|Y_{t-k})
\end{equation}

\noindent where $k$ and $d$ are positive values representing possibly different time points in the past depending on the processes. Indeed, they can represent vectors denoting multiple time-points, in order to study the transfer over a range of delays. 

Reformulating transfer entropy terms as partial information atoms, \cite{williams2011generalized} showed that transfer entropy can be measured as the sum of unique information from $X_{t-d}$ about $Y_t$ that did not come from $Y_{t-k}$, and the synergistic information that is available due to the combined knowledge of $X_{t-d}$ and $Y_{t-k}$. 

\begin{equation}
    TE_{X\rightarrow Y} = U(Y_t;X_{t-d}) + S(Y_t;{X_{t-d},Y{t-k}})
\end{equation}

\end{document}